\documentclass{amsart}

\usepackage{epsfig}

\newtheorem{theorem}{Theorem}
\newtheorem{lemma}[theorem]{Lemma}

\usepackage{geometry} 
\usepackage{amsmath}
\usepackage{graphicx}
\usepackage{epsfig}
\usepackage{latexsym}
\usepackage{epsfig}
\newcommand{\PP}{{\mathbb P}}

\newcommand{\Sc}{\mathcal S}
\begin{document}

\title[Phylogenetic decisiveness]{Characterizing phylogenetically decisive taxon coverage}
\author{Mike Steel and Michael J. Sanderson}

\thanks{We thank Michelle M. McMahon for discussion, and the  Alexander von Humboldt Foundation and the US NSF  and the for research support}

\address{Mike Steel:  Allan Wilson Centre for Molecular Ecology and Evolution, Department of Mathematics and
  Statistics, University of Canterbury, Christchurch, New Zealand; 
  Michael Sanderson:  Department of Ecology and Evolutionary Biology, University of Arizona, Tucson, AZ 85721}

\email{m.steel@math.canterbury.ac.nz}

\subjclass{05C05; 92D15}

\begin{abstract}
Increasingly,  biologists are constructing evolutionary trees on large numbers of overlapping sets of taxa, and then combining them into a `supertree' that classifies all the taxa. In this paper, we
ask how much coverage of the total set of taxa is required by these subsets in order to ensure we have enough information to reconstruct the supertree uniquely. We describe two results - a combinatorial characterization of the covering subsets to ensure that at most one supertree can be constructed from the smaller trees (whatever trees these may be) 
 and a more liberal analysis that asks only that the supertree is highly likely to be uniquely specified by the tree structure on the covering subsets.
\end{abstract}

\maketitle

\section{Introduction}

The scale of phylogenetic analysis has been growing steadily both in the number of taxa and the number of loci. 
Data from different loci are combined either directly into a single inference or indirectly by first building trees from each locus and combining trees as a ``supertree''. 
 Regardless of approach, large-scale phylogenetic data sets derived from genome resources \cite{dun} or mining databases like GenBank \cite{gol, mcm, smi} tend to exhibit a high proportion of missing entries (taxa missing from taxa sets for different loci or input trees) -- 55\% to 96\% in the papers just cited. Wiens \cite{wie} has argued that the effect of missing data on the accuracy of tree inference is minimal as long as these missing data are randomly distributed and counterbalanced by enough data overall. 
 
 However, the pattern of missing entries is highly nonrandom, especially in the data mining studies, as the pattern is  determined by numerous sample biases in the databases (for examples, see the PhyLoTA Browser database \cite{phylota}). Moreover, few analytic results are available to complement simulation based studies of this problem.
 
  In this paper, we mathematically  address the question of whether a given collection of subsets of taxa would suffice to reconstruct a tree uniquely  for all the taxa, if we can infer a tree correctly on each of the subsets.   Our study is also motivated by some recent mathematical work concerning supertree construction under various taxon coverage conditions,
 \cite{gro, gro2, ste2}.

\subsection{Definitions}
We begin by recalling some basic definitions from phylogenetic theory. Following \cite{sem}, given a set $X$ of taxa, a
 {\em binary phylogenetic $X$--tree} is a tree $T$ in which the degree 1 vertices (leaves of $T$) consist of the set $X$ and all the remaining vertices of $T$ are unlabelled and of degree $3$.
Fig. 1 shows two of the 15 distinct binary phylogenetic $X$--trees for $X=\{1,2,3,4,5\}$.  For a binary phylogenetic tree $T$ and a subset $Y$ of $X$, let $T|Y$ denote the induced 
binary phylogenetic tree on leaf set $Y$ (the tree obtained from the minimal subtree connecting $Y$ by suppressing any vertices of degree $2$). 
 A {\em quartet tree} is a binary phylogenetic tree on four leaves. For such a tree, with leaves $a,b,c,d$, we write $ab|cd$ if the interior edge of the tree
separates the pair $a,b$ from $c,d$.

Let $\Sc$ be a collection of subsets of a set $X$, and let $n = |X|$ throughout.  
We say that $\Sc$ is {\em phylogenetically decisive} if it satisfies the following property: If $T$ and $T'$ are binary phylogenetic $X$--trees, with
$T|Y = T'|Y$ for all $Y \in \Sc$, then $T = T'$.  In other words, for any binary phylogenetic $X$--tree $T$, the collection of induced subtrees $\{T|Y: Y \in \Sc\}$ uniquely determines
$T$ (up to isomorphism).   

Let $Q_\Sc$ be the set of all quartets from $X$ that lie in at least one set in $S$.  That is:
$$Q_\Sc: = \bigcup_{Y \in S} \binom{Y}{4}.$$
Note that $\Sc$ is phylogenetically decisive if and only if $Q_\Sc$ is phylogenetically decisive since 
$T|Y = T'|Y$ if and only if $T|q = T'|q$ for all $q \in \binom{Y}{4}$ \cite{ste}.
It is easily shown that if $\Sc$ is phylogenetically decisive then: 
$$\binom{X}{3} \subseteq \bigcup_{Y \in \Sc} \binom{Y}{3}.$$
In other words, all three--taxon subsets of $X$ must be present as a subset of some element $Y$ of $\Sc$ (this is Lemma 6.2.1 of \cite{hum}).  However, this 
necessary condition for phylogenetic decisiveness  can be shown to be insufficient (an example is provided in \cite{hum}).  One sufficient condition has been known since 1992 \cite{ste}; namely if
$Q_{\Sc}$ contains all quartets of the form $\{x_0, x,y,z\}$ for some fixed $x_0 \in X$, and all distinct $x,y,z \in X-\{x_0\}$, then
$\Sc$ is phylogenetically decisive. However, this sufficient condition is not necessary, as the following example shows.

\subsection{Example}

Let $\Sc =\{\{1,2,3,4\},\{1,2,3,5\}, \{2,3,4,5\}, \{1,3,4,5\}\}$. Then $\Sc$ is a phylogenetically decisive collection of subsets of $X= \{1,2,3,4,5,\}$, 
that is, each of the 15 binary phylogenetic $X$--trees is determined by
the collection $T|Y$ for $Y \in \Sc$. For example, for the two trees $T, T'$ in Fig. 1., we have different induced quartet trees  $T|Y = 12|34, T'|Y = 13|24$  by selecting the taxon set $Y=\{1,2,3,4\}$ from $\Sc$..  Notice that in this example,  no element of $X$ lies in every set in $\Sc$. Theorem~\ref{mainthm} below will allow us to easily verify that $\Sc$ is phylogenetically decisive.

\begin{figure}[ht] \begin{center}
\resizebox{11cm}{!}{
\input{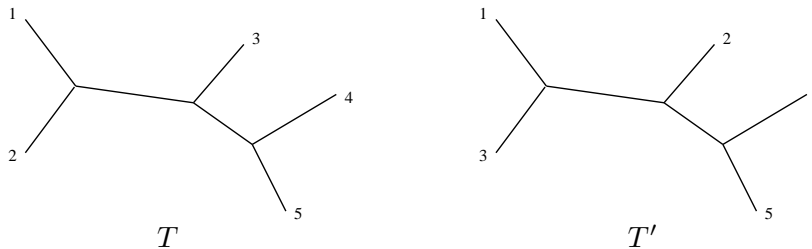}
} \caption{Two binary phylogenetic $X$--trees for $X=\{1,2,3,4,5\}$.  If we take $Y= \{1,2,3,4\}$ then $T|Y = 12|34$ and $T'|Y = 13|24$.}
\end{center}
\label{figure1}
\end{figure}

\section{Characterizing decisiveness}

Our first result provides  a purely combinatorial characterization for phylogenetic decisiveness, and this is provided as follows. We say that a collection $\Sc$ of subsets of $X$ satisfies the 
{\em four--way partition property (for $X$)} if, for all  partitions of $X$ into four (disjoint) sets $A_1, A_2, A_3, A_4$ (with $A_1\cup A_2 \cup A_3 \cup A_4= X)$
there exists $a_i \in A_i$ for $i=1,2,3,4$ for which $\{a_1, a_2, a_3, a_4\} \in Q_{\Sc}.$
We begin with a useful lemma.   Recall that a {\em cherry} of a tree  is a pair of leaves that are adjacent to the same vertex. 

\begin{lemma}
Suppose that $T$ is a binary phylogenetic $X$--tree, and $a,b \in X$. 
\begin{itemize}
\item[(i)] If $a,b$ forms a cherry of $T$, then for every pair of subsets $C, D$ that partition $X-\{a,b\}$, and every $c \in C, d\in D$, we have $T|\{a,b,c,d\} = ab|cd$.
\item [(ii)]Conversely, if $a, b$ does not form a cherry of $T$, then there exists a pair of subsets $C, D$ that partition $X-\{a,b\}$, such that, for every $c \in C$ and every $d \in D$, $T|q$ is different to $ab|cd$.
\end{itemize}
\end{lemma}

{\em Proof:} Part (i) of the claim is clear. For part (ii), we show that if $a,b$ is not a cherry of $T$ then we can construct a partition of $X-\{a,b\}$ that satisfies the property described.  Consider the path in $T$ connecting $a$ and $b$.  If $a,b$ is not a cherry, this path has at least two trees hanging off it. Let $C$ be the leaf set of the hanging tree that is closest to $a$, and let $D = X-C-\{a,b\}$.  Then, regardless of which 
element $c$ we select in $C$ and which element $d$ we select in $D$, we have $T|\{a,b,c,d\} = ac|bd$.  
\hfill$\Box$

\begin{theorem}
\label{mainthm}
A collection $\Sc$ of subsets of $X$ is  phylogenetically decisive if and only if $\Sc$ satisfies the four--way partition property  for $X$.
\end{theorem}
{\em Proof:}
We first show that the condition is necessary.  Suppose, to the contrary, that a four--way partition of $X$ exists as described, but without a  quartet $\{a_1, a_2, a_3, a_4\} \in Q_{\Sc}$ with
$a_i \in A_i$.      Let $T$ be any binary phylogenetic tree, obtained by taking arbitrary binary rooted phylogenetic trees on leaf sets $A_1, A_2, A_3, A_4$, and identifying the roots of these four
trees with the leaves of a quartet tree. Let $T'$ be one of the two trees obtained from $T$ by performing a nearest neighbor interchange about the central edge of the quartet (to which the four
rooted trees were attached).  The only quartets from $X$ that $T'$ resolves differently from $T$ are quartets that contain one leaf from each of the sets $A_1, A_2, A_3, A_4$ and we have assumed there
is no such quartet in $Q_{\Sc}$. This shows that $\Sc$ is not phylogenetically decisive.

We next show that the condition is sufficient.  Suppose that $T$ is any binary phylogenetic $X$-- tree. We must show that no other binary tree displays the collection of quartet trees
$Q_T:= \{T|q: q \in Q_\Sc\}$. We will use induction on $|X|$.  The result clearly holds for $|X|=4$.  Now suppose $T$ has $n>4$ leaves and that $T'$ is a phylogenetic tree that displays $Q_T$.
We will show that $T' = T$.

 Lemma 1 allows us to use $Q_\Sc$  to identify when a pair $a,b$ is a
cherry of $T$. The argument is as follows. For each pair $a,b$, consider all choices of $C,D$ that partition $X-\{a,b\}$. By our assumption concerning $\Sc$ (taking $A_1 = \{a\}, A_2 =\{b\}, A_3 = C, A_4=D$),  it follows that there exists $c\in C, d\in D$ such that $\{a,b,c,d\} \in Q_\Sc$, and so some resolution of $a,b,c,d$ is in $Q_T$. If this resolution is different from $ab|cd$ then we discard $a,b$ as a candidate for being a cherry of any tree that displays $Q_T$, including $T$ and $T'$ (by  Lemma 1(i)).  On the other hand, if for every choice of $C,D$, we have the resolution $ab|cd$ in $Q_T$ then $a,b$ must be a cherry of every tree that displays $Q_T$, including $T$ and $T'$ (by Lemma 1(ii)).

Now, consider the set $X'$ obtained from $X$ by deleting $b$, and let $\Sc'$ be the collection of subsets of $X$ obtained from $\Sc$ by replacing each occurrence of $b$ in $Y \in \Sc$ by $a$ (if $a, b$ appear together in some set $Y$, then we simply delete $b$ from that set). We claim that if $\Sc$ satisfies the four--way partition property for $X$ then $\Sc'$  satisfies this property  for $X'$.  
Consider a partition $A_1, A_2, A_3, A_4$ of $X'$.  The element $a$ lies in one of these sets - let us say $A_1$. Consider the four--way partition of $X$ given by:
$A_1 \cup\{b\}, A_2, A_3, A_4.$  Then, by assumption, there exists $a_1 \in A_1 \cup \{b\}, a_i \in A_i (i=2,3,4)$ with  $\{a_1, a_2, a_3,a_4\} \in Q_\Sc.$  Now, $b$ is not one of $a_2, a_3, a_4$, and so,  regardless of
whether $a_1$ is $a$, or $b$ or neither, we have $\{a_1, a_2, a_3,a_4\} \in Q_\Sc'$  (Note that if $Y$ is a set in $\Sc$ of size $4$ containing $a,b$ then this set will not produce a quartet in $Q_\Sc'$,  since on deleting $b$, we obtain a set of size 3 -- however  this does not create a problem since we have at least three elements of $X-\{a,b\}$ in $\{a_1, a_2, a_3, a_4\}$). 

Let $T_b = T|X'$ be the  binary phylogenetic $X'$--tree obtained from $T$ by deleting leaf $b$ (and its incident edge), and let 
$Q_{T_b}=\{T_b|q: q\in Q_{\Sc'}\}$.    By induction (noting that $|X'|<|X|$ and that $\Sc'$ satisfies the four--way partition property for $X'$), $T_b$ is the unique tree that displays $Q_{T_b}$.  However, the tree $T'_b$ obtained from $T'$ by deleting leaf $b$ also displays $Q_{T_b}$ since $T'$ displays $Q_{\Sc}$ and $a,b$ is a cherry of $T'$.  Thus $T'_b = T_b$ and thus, $T'=T$ as required
\hfill$\Box$

\subsection{Remarks}
\begin{itemize}
\item
The fact that if $\Sc$ is phylogenetically decisive then every element subset $\{a,b,c\}$ of $X$ must be contained in a quartet within  some set $Y \in \Sc$ follows immediately from Theorem~\ref{mainthm} by  taking $A=\{a\}, B = \{b\}, C=\{c\}$ and $D=X-\{a,b,c\}.$
\item
The argument in the proof suggests an algorithm for building a tree based on identifying a cherry and recursion. 
\item
The computational complexity of determining whether an arbitrary collection of $\Sc$ of subsets of $X$ is phylogenetically decisive seems an interesting question, since the number of
all four--way partitions is exponential. For practical applications, a simple but fast measure for quantifying the degree of  phylogenetic  decisiveness of a set $\Sc$ would be to generate uniformly at random a large number of four-way partitions of $X$ and ask for what proportion of the resulting 4-way partitions  ($A_1, A_2, A_3, A_4$) there exists $a_i \in A_i$ for $i =1,\ldots, 4$ with $\{a_1,a_2, a_3, a_4\} \in Q_{\Sc}$.   
If this proportion is strictly positive then $\Sc$ is not phylogenetically decisive but it may still be of interest to know how `close' to phylogenetically  decisive it is by this measure. 
\item
The concept of phylogenetic decisiveness is related to, but different from, the weaker concept of a phylogenetic `grove' from \cite{gro, gro2}.
\end{itemize}

\section{Decisive sets for random trees}

The combinatorial condition for phylogenetic decisiveness is very strong, and in this section we describe a condition that reflects the fact although all trees might not be determined by how they resolve
certain sets of taxa that cover $X$, nearly all trees will be.  For a collection $\Sc =\{Y_1, \ldots, Y_k\}$, we say that
{\em $\Sc$ is decisive for a tree $T$}  provided $T$ is the only tree that displays $T|Y_1, \ldots, T|Y_k$.  
Thus if $\Sc$ is decisive then it is decisive for every tree $T$, but the converse is
certainly not true - for instance, for every binary phylogenetic tree $T$, there is a set of just $n-3$ quartets for which $\Sc$ is decisive for $T$ \cite{ste}.  For example, in Fig. 1, the set
$\{\{1,2,3,4\}, \{1,3,4,5\}\}$  is decisive for $T$ but not for $T'$.

By a {\em random tree}, we mean a binary phylogenetic $X$--tree chosen uniformly at random from the set of $(2n-5)!!$  binary phylogenetic $X$--trees. 
  Thus we can talk about the probability that a given $\Sc$ is decisive for a random tree (it is simply the proportion of binary phylogenetic $X$--trees for which $\Sc$ is decisive).  Note that if $\Sc$ consists of two sets $Y_1, Y_2$ and each of these sets contains a taxon that is not in the intersection $Y_1 \cap Y_2$
then $\Sc$ is not decisive, even if just one taxon is unique to $Y_1$ and to $Y_2$ (provided the intersection contains at least two elements).  However, the following result shows that $\Sc$ is very likely to be decisive for a random tree, even when  several (but not too many) taxa lie outside the intersection.  

\begin{theorem}
\label{second}
Let $\Sc = \{Y_1, Y_2\}$ where $k= |Y_1 \cap Y_2|$ and $\lambda_1 = |Y_1-Y_1 \cap Y_2|, \lambda_2 = |Y_2 - Y_1 \cap Y_2|.$  
Let $p:= \PP[\Sc \mbox{ is decisive for  a random tree } T].$
Then: 
\begin{itemize}
\item[(i)] $p  \geq  1- \frac{3\lambda_1\lambda_2}{(2k-3)};$
\item[(ii)] $p \leq  \exp\left(-\frac{\lambda_1\lambda_2}{\lambda_1+\lambda_2+k- 5/2}\right).$
\end{itemize}
In particular, if $\lambda_1\lambda_2= o(k)$ then $p=1 - o(1)$.
\end{theorem}

{\em Proof:}
First observe that if $\lambda_1=0$ or $\lambda_2=0$ then $p=1$, and both (i) and (ii) apply (as tight bounds), so we will henceforth assume that $\lambda_1, \lambda_2 >0$.

Let $Y: = Y_1 \cap Y_2$, and consider $T|Y$.  If $T$ is a random tree with leaf set $X$, then $T|Y$ is a random tree with leaf set $Y$.  Moreover,  we can generate a binary phylogenetic $X$--tree uniformly at random by the following randomized leaf-attachment process \cite{pen}.  Take a given ordering of the taxa (note that this ordering is not necessarily selected randomly). We construct a sequence of trees
beginning with a tree consisting of the first two taxa in the ordering, connected by an edge, and ending with the tree $T$.  The process of constructing the next tree in the sequence from the previous is as follows: Select one of the edges of the tree so far constructed uniformly at random, subdivide this edge and make the midpoint adjacent to a new leaf (via a new edge) that is labelled by the next taxon in the ordering that has not appeared in the tree so far constructed. 

With reference to $Y, Y_1, Y_2$, we will select an ordering where the taxa in $Y$ come first, then those in $Y_1-Y$ and finally those in $Y_2-Y$ (any such ordering satisfying this constraint is adequate) to obtain
a random binary tree (with uniform probability) on  leaf set $X: = Y_1 \cup Y_2$.  Note that each element of $y$ of $Y_1-Y$ or of $Y_2-Y$ has a unique nearest edge $e$ of $T|Y$, which we will denote by
$e(y)$. Moreover, the condition for $T$ to be the only tree
that displays the induced trees  $T|Y_1$ and $T|Y_1$ is that the sets of edges $E_1: = \{e(y): y \in Y_1 -Y\}$ and $E_2= \{e(y): y \in Y_2-Y\}$ are disjoint subsets of the total set of edges of $T|Y$
(by Theorem 1 of \cite{boc}).  Conditional on $T|Y$ and $E_1$, consider the probability $p' = p'(T|Y, E_1)$ of the event that the leaf attachment of the leaves in $Y_2-Y$ results in $E_2$ being disjoint from $E_1$ (and, as noted, this event implies  that $T$ is the only tree that displays $T|Y_1$ and $T|Y_2)$. We have: 
\begin{equation}
\label{usefuleq}
p' = \frac{x}{(\lambda+2\lambda_1)} \cdot \frac{(x+2)}{(\lambda+2\lambda_1+2)} \cdots \frac{(x+2(\lambda_2-1))}{(\lambda+2\lambda_1 + 2(\lambda_2 - 1))},
\end{equation}
where $\lambda  = 2k-3$ is the number of edges of $T|Y$, and $x = \lambda - |\{e(y): y \in E_1\}|$. 

{\em Proof of (i):}
From Eqn. (\ref{usefuleq}) we have:
\begin{equation}
\label{usefuleq2}
p'  \geq   \left(\frac{x}{\lambda+2\lambda_1}\right)^{\lambda_2}
\end{equation}
Now,  the smallest possible value of the right-hand side term in (\ref{usefuleq2}) over all choices of $T|Y, E_1$, is realized when  $x$ takes its smallest possible value -- or, equivalently, when $E_1$ takes its maximal possible value of $\lambda_1$ (i.e. $e(y)$ is a different edge of $T|Y$ for each $y \in Y_1-Y$), in which case $x = \lambda-\lambda_1$.  Substituting this into (\ref{usefuleq}) gives:
$$p' \geq \left(\frac{\lambda-\lambda_1}{\lambda+2\lambda_1}\right)^{\lambda_2} = \left(\frac{1-\lambda_1/\lambda}{1+2\lambda_1/\lambda}\right)^{\lambda_2} \geq \left(1-\frac{3\lambda_1}{\lambda}\right)^{\lambda_2} \geq 1-3\lambda_1\lambda_2/\lambda.$$
This lower bound is conditional on the two random variables $T|Y$ and $E_1$; however, it depends only on these only via the quantities  $\lambda$ and $\lambda_1, \lambda_2$ which are fixed in advance, and so the bound applies also 
without conditioning.  This completes the proof of (i).

{\em Proof of (ii):}   
From Eqn. (\ref{usefuleq}), we have:
\begin{equation}
\label{usefuleq3}
p'  \leq   \left(\frac{x+2\lambda_2-2}{\lambda+2\lambda_1+2\lambda_2-2}\right)^{\lambda_2}
\end{equation}
and since $x \leq \lambda-1<\lambda$, we have:
$$p' < \left(1-\frac{2\lambda_1}{\lambda+2\lambda_1+2\lambda_2-2}\right)^{\lambda_2} \leq \exp\left(-\frac{\lambda_1\lambda_2}{\lambda_1+\lambda_2+\lambda/2-1}\right),$$
from which (ii) now follows for similar reasons to the conclusion of the proof of part (i). 

\hfill$\Box$

\subsection{Remark}
Using the theory of Polya Urn models, one could, in principle, obtain an exact but complex expression for $p$ in Theorem~\ref{second}.  However, a more interesting problem would
be to obtain bounds for the probability that $\Sc$ is decisive for a random tree, when $\Sc$ consists of more than two sets.

\end{document}